\documentclass{epl}

\usepackage{subfigure}

\title{Influence of hydrodynamics on many-particle diffusion in
       2D colloidal suspensions}
\author{E. Falck\inst{1,2} \and J. M. Lahtinen\inst{1} \and
        I. Vattulainen\inst{1,2}\and T. Ala-Nissila\inst{1,3}}
\institute{
  \inst{1} Laboratory of Physics - Helsinki University of Technology,
           P.O. Box 1100, FIN--02015 HUT, Finland\\
  \inst{2} Helsinki Institute of Physics - Helsinki University of
           Technology, P.O. Box 1100, FIN--02015 HUT, Finland\\
  \inst{3} Department of Physics - Box 1843, Brown University,
           Providence R.I. 02912--1843, U.S.A.
}

\shorttitle{Influence of hydrodynamics on many-particle etc.}
\shortauthor{E. Falck {\it et al.}}

\begin{document}

\maketitle

\begin{abstract}
We study many-particle diffusion in 2D colloidal suspensions with
full hydrodynamic interactions through a novel mesoscopic
simulation technique. We focus on the behaviour of the effective
scaled tracer and collective diffusion coefficients $D_T(\rho) /
D_0$ and $D_C(\rho) / D_0$, where $D_0$ is the single-particle
diffusion coefficient, as a function of the
density of the colloids $\rho$. At low Schmidt numbers 
$Sc={\cal O}(1)$, we find that hydrodynamics has essentially no effect 
on the behaviour of $D_T(\rho)/D_0$. At larger $Sc$, $D_T(\rho)/D_0$
is enhanced at all densities, although the differences compared to
the case without hydrodynamics are minor. The collective 
diffusion coefficient, on the other hand, is much more
strongly coupled to hydrodynamical conservation laws and is
distinctly different from the purely dissipative case.
\end{abstract}

{\it Introduction} -- The dynamics of Brownian particles in
confined geometries, and in two dimensions (2D) in particular, is
an important theoretical problem with applications in surface
science and colloidal systems \cite{Nag96}. Examples of
fundamental questions that have been addressed recently are the
form of effective interactions between macroions in a colloidal
suspension \cite{Acu98,Bru02} and the effects of hydrodynamic
interactions (HIs) on the diffusive properties of colloidal particles
\cite{Zah97,Rin99,Pes00,Kol02}.

So far, most studies have focused on {\it self-diffusion}
of particles in 2D. In the ideal case of no external potential and
without HIs, the density dependent
self-diffusion coefficients of 2D hard disk particles have been
recently determined using numerical simulations
\cite{lowen96,lahtinen01} and various theoretical approximations
\cite{lahtinen01,hess83,indrani94}. While the
single-particle limit in the ideal case is trivial, no exact
analytic results exist for finite densities. In this regime
complicated many-body effects manifest themselves
through memory effects in the motion of tagged colloidal particles.

The situation is even more complicated when the HIs 
mediated by the solvent in a colloidal suspension
are taken into account. Recent work on the self-diffusion of
colloidal particles in 2D and quasi-2D
\cite{Zah97,Rin99,Pes00,Kol02} indicate that HIs do indeed
influence self-diffusion. In many cases it has been found that
diffusion is enhanced \cite{Zah97,Rin99,Pes00,Kol02} by the
HIs. Nevertheless, it is possible that the nature and magnitude of
these subtle effects in a given system depend on the relative
importance of the HIs and other interactions.

While the self-diffusion properties of 2D colloidal systems
are relatively well understood, much less is known about 
{\it collective diffusion} in 2D colloidal suspensions. Nevertheless,
collective diffusion plays a crucial role in processes such
as spreading and phase separation, as it describes the decay rate
of density fluctuations in a system. While the case without HIs
has been considered recently \cite{lahtinen01,Acu00}, the
theoretical understanding of 2D situations with HIs 
is very limited. This is, in part, due to theoretical
difficulties when dealing with collective transport in 2D
liquids. Furthermore, realistic numerical simulations of
collective diffusion in two-phase colloidal systems with full
hydrodynamics have turned out to be a considerable methodological challenge.

In this letter, our purpose is to employ a recently proposed
mesoscopic simulation method \cite{malevanets99,malevanets00a} to
shed light on some of the fundamental issues of many-particle
diffusion in 2D colloids. To this end, we consider an ensemble of
colloidal particles that interact mutually with a short-range
repulsive potential and long-range HIs. The
main issue we want to address here is the influence of
hydrodynamics in the diffusive dynamics of this system, as the
extent of HIs is varied in a controlled fashion by tuning the
Schmidt number of the system (see below). In particular, we are
interested in the behaviour of the effective collective diffusion
coefficient as a function of the density of the colloidal
particles and the impact of the HIs on this dependence.

{\it Schmidt number} --
An important quantity measuring the properties of a fluid in
equilibrium is the dimensionless Schmidt number $Sc$, defined
as the ratio of momentum diffusivity to mass diffusivity:
\begin{equation}
Sc=\frac{\nu}{D}.
\label{EQschmidt}
\end{equation}
Here $\nu = \eta / \rho_s$ is the kinematic viscosity of the
fluid, $\eta$ being the viscosity, $\rho_s$ the density of the
fluid and $D$ the tracer diffusion coefficient of the fluid
particles. In a real fluid such as water $Sc = {\cal O}(10^3)$.
Theoretical arguments, too, often include the assumption that
hydrodynamic fluctuations have reached a steady state on the
timescale of the motion of the colloidal particles. The situation
can be quite different in computer simulations. In molecular
dynamics (\acro{MD}) simulations the Schmidt number is typically large.
Unfortunately, \acro{MD} is computationally too intensive for simulations
of colloidal suspensions with explicit solvent, especially at high
densities. Mesoscopic simulation techniques such as dissipative
particle dynamics (\acro{DPD}) reduce the computational
cost. On the other hand, in recent \acro{DPD} simulations $Sc = {\cal O}(1)$
\cite{warren97}. It is not quite clear how a Schmidt number
several magnitudes smaller than those measured in realistic
systems influences the properties of {\it e.g.} colloidal
suspensions. In our opinion, the understanding of the
influence of $Sc$ on colloidal dynamics is necessary. 
It is also crucial to develop new approaches to bridge
the various scales in colloidal systems and, consequently, to
allow a direct comparison between physical model studies and
experiments.

{\it Model and methods} --
The model system we consider here comprises an ensemble of disks
immersed in a 2D liquid. The dynamics of the system is simulated
using a novel mesoscopic technique introduced by Malevanets and
Kapral (\acro{MK}) \cite{malevanets99,malevanets00a}. The \acro{MK} method is
essentially a hybrid \acro{MD} model, where the
colloid is treated microscopically and the solvent obeys
coarse-grained dynamics. The interaction potential between two
solvent particles is zero, while the solvent-colloid and
colloid-colloid interactions can be chosen at will. For the
purposes of the coarse-grained solvent dynamics, the time is
partitioned into segments $\tau$, and the system itself is divided
into so-called collision volumes or cells. During one collision
step $\tau$, the system obeys Newton's equations of motion, {\it
i.e.} the time evolution is taken care of by \acro{MD}. The effective
interactions between two solvent molecules occur at each $\tau$:
a collision event takes place, and the velocities
of the solvent particles are transformed according to
$\mathbf{v}_i(t+\tau) = \mathbf{V}+\bm{\omega}
\cdot[\mathbf{v}_i(t)-\mathbf{V}]$.
Here $\mathbf{v}_i$ is the velocity of particle $i$, $\mathbf{V}$ is the
average velocity of all the particles in the cell the
particle $i$ belongs to and $\bm{\omega}$ is a random rotation
matrix over an angle $\alpha$ chosen separately for each cell. It
can be shown \cite{malevanets99} that this multiparticle collision
dynamics conserves the momentum and energy in each collision
volume. Furthermore, it gives a correct description of the
hydrodynamics of the velocity field. The method also reduces
computing times significantly, as the tedious calculation of
solvent-solvent forces can be omitted.

\begin{figure}[hbt]
\centering
\mbox{\subfigure
{\includegraphics[scale=0.33,clip=]{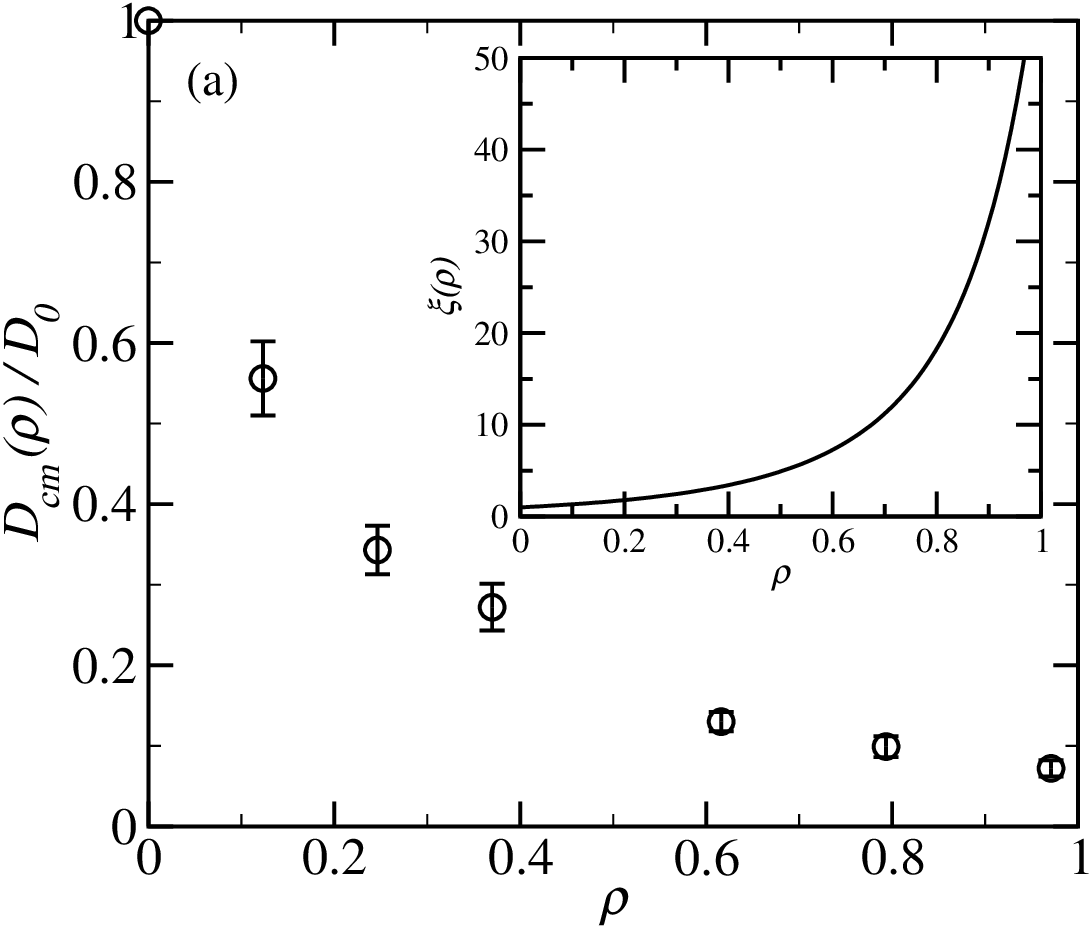}}\quad
\subfigure
{\includegraphics[scale=0.33,clip=]{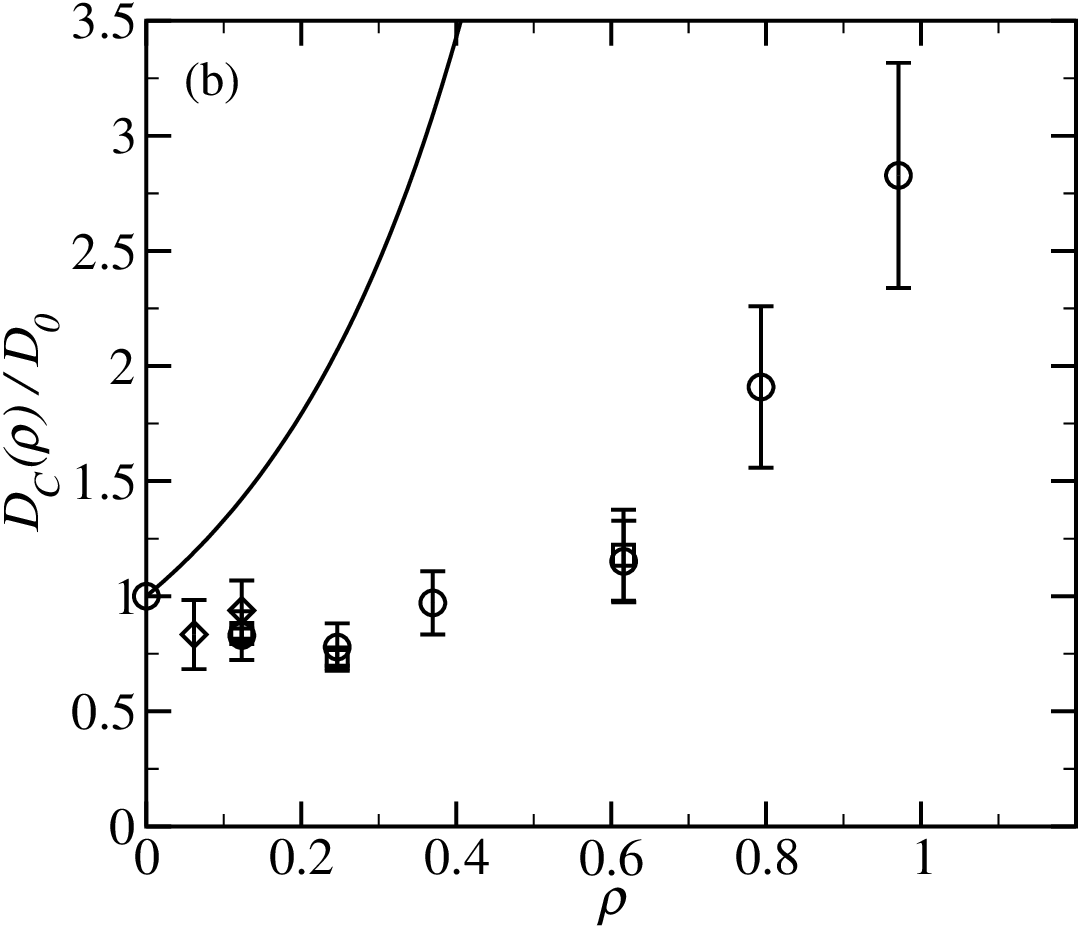}}}
\caption{\label{FIGcollective}(a) Scaled CM mobility $D_{cm}(\rho)/D_0$ as
a function of $\rho$ for $Sc \approx 1$.
The inset shows the thermodynamic factor $\xi$ from the Boublik
approximation \cite{boublik75}. (b) Scaled effective collective 
diffusion coefficients
$D_C(\rho) / D_0$ as a function of $\rho$. Open circles are
results with $Sc \approx 1$, obtained from simulations without
colloid-solvent interactions. Open diamonds are results from
simulations where the colloid-solvent interactions are present.
Squares, in turn, are from computations where $Sc \approx 100$.
For reference, the solid line is $D_C(\rho) / D_0$ in the case
where HIs have not been taken into
account \protect\cite{lahtinen01}.}
\end{figure}

The interactions between the colloid particles are strongly
repulsive and of short range. The interaction potentials are
of the form
\begin{equation}
V(r) = \left\{ \begin{array}{ll}
\epsilon \left( \frac{\sigma}{r} \right)^n, &
          \mbox{$r \leq r_c \equiv 2.5 \sigma$}; \\
0, &\mbox{$r > r_c$},
\end{array}
\right.
\label{EQpotential}
\end{equation}
where $r$ is the separation between two particles. Here we set
$n=12$, which allows a direct comparison with previous
calculations on a smooth surface without hydrodynamics
\cite{lahtinen01}. The colloid-colloid interaction parameters were
chosen to be $\sigma_{cc} = 2 \sigma$ and $\epsilon_{cc} =
\epsilon$, while the colloid-solvent parameters are given by
$\sigma_{cs} = \sigma$ or zero\footnote{ In the case where the
direct solvent-colloid interaction is absent, the colloidal
particles participate in the collision events
\cite{malevanets00b}. In other words, the solvent-colloid
interaction is described indirectly through collisions. As a
consequence the calculations speed up significantly. We have
verified explicitly that the results obtained from simulations
without solvent-colloid interactions are in quantitative agreement
with those computed including the interaction directly, see fig.\
\ref{FIGcollective}.(b)} and $\epsilon_{cc} = \epsilon$. The colloidal
particles have a mass of $m_c = 5 m$ and the solvent mass is set
to $m_s = m$ or $m_s = 0.5 m$. The parameters $\sigma$,
$\epsilon$ and $m$ now define our system of units, and hence our
unit of time is given by $\tau_{LJ} = \sigma (m /
\epsilon)^{1/2}$.

The simulations were performed at a temperature
$k_B T = 2$ and for the solvent density we used $\rho_s = 1$.
\footnote{The solvent and colloidal densities
have been defined as $\rho_s \equiv N_s / (A / \sigma^2)$ and
$\rho \equiv N / (A / \sigma_{cc}^2)$, respectively. 
Here $A$ is the area of the
system, $N_s$ the number of solvent particles and $N$ the number
of colloidal particles.}
The equations of motion were integrated using the velocity
Verlet algorithm with a time step \,$\Delta t = 0.005$. The
linear size of our simulation box was set to either $L = 200$
or $L = 100$. Periodic boundary conditions were employed in
all simulations. Within our statistical accuracy, the scaled diffusion
coefficients did not change with system size. However, we did not carry 
out a systematic finite-size scaling analysis.

The choice of the parameters that determine the collision dynamics
fixes the properties of the coarse grained solvent, and in
particular the Schmidt number. The value of $Sc$ is affected
{\it e.g.} by the choice of the collision step $\tau$,
the linear size of the collision volume $\ell_c$ and the
collision rule, {\it i.e.} the choice of the random rotation
matrix $\bm{\omega}$. We have used different sets of parameters in
the simulations. These sets lead to different Schmidt numbers
and kinematic viscosities. 
A list of the three different parameter sets is shown in Table \ref{TABparam}.
\begin{table}[t]
\begin{center}
\begin{tabular}{|l|l|l|l|l|l|l|r|} \hline\hline
{\bf$Sc$} & {\bf$\nu \, [\sigma^2 \tau_{LJ}^{-1}]$} &
 {\bf$\tau \, [\tau_{LJ}]$}  &
{\bf$\ell_c \, [\sigma]$} & {\bf$\alpha$}
& {\bf$m_s \, [m]$}  & {\bf$L \, [\sigma]$}  \\ \hline
$1$       & $0.82(1)$       & $0.5$  & $2$   & $\pm 90 ^{\circ}$ &
 $1$       & $200$ \\
$20$      & $3.70(2)$       & $0.1$  & $2$   & $\pm 125^{\circ}$ &
 $0.5$     & $100$ \\
$100$     & $9.11(2)$       & $0.1$  & $2$   & $\pm 170^{\circ}$ &
 $1$       & $100$ \\ \hline\hline
\end{tabular}
\end{center}
\caption{\label{TABparam} A summary of the three different
parameter sets used. See text for details.}
\end{table}
The systems with $Sc > 1$ have a mean free path (distance
traveled during $\tau$) of the order of $\lambda \approx 0.1
\ell_c$. In this case, to avoid any unphysical correlations at
short time and length scales, we have employed the random
grid-shifting procedure proposed in ref. \cite{ihle01}.

{\it Results and discussion} --
Perhaps the simplest transport coefficient is the tracer
diffusion coefficient $D_T$, which describes the motion 
of a tagged tracer
particle as $D_T = \lim_{t \to \infty} D_T(t)$, where
\begin{equation}
D_T(t) =  \frac{1}{d} \int_0^{t} \mathrm{d}t' \langle
\frac{1}{N}\sum_{i=1}^N\mathbf{v}_i(t')\cdot\mathbf{v}_i(0) \rangle,
\label{EQtracer}
\end{equation}
among $N$ identical particles. Here $d=2$ is the dimension of
the system and $\mathbf{v}_i(t)$ is the velocity of particle $i$
at time $t$. The quantity $\phi(t) \equiv \langle
\mathbf{v}_i(t)\cdot\mathbf{v}_i(0) \rangle$ is the velocity
autocorrelation function. In evaluating the diffusion coefficients
we employ the memory expansion method \cite{ying98}.

For colloidal systems with HIs, it has been known since the 1970s
that there are long-time tails in the velocity autocorrelation
functions. These have been observed in \acro{MD}
\cite{alder70} and lattice-Boltzmann \cite{vdhoef91} computer
simulations, and will lead to a divergence of the tracer diffusion
coefficient in 2D \cite{alder70,dorfman72}. In this
case, the diffusion coefficients $D_T(t)$ can be considered to be
time-dependent, effective quantities. When tracer diffusion in
concentrated suspensions is investigated experimentally, it is
common to single out the effects of the HIs by concentrating on the
so-called short-time diffusion coefficients
\cite{medina88,Zah97,Rin99,Kol02}. These are measured at times
much shorter than $\tau_I$, the time it takes the tracer particles
to diffuse the average distance between the suspended colloidal
particles. For this definition to be meaningful, such times should
be significantly larger than the time $\tau_B$ it takes for the
velocities of the colloidal particles to relax. In our case, for
concentrated solutions and small values of $Sc$ in particular, the
time interval between $\tau_B$ and $\tau_I$ becomes very narrow.
Hence, this definition for short-time diffusion coefficients
cannot be used.

\begin{figure}[hbt]
\centering
\mbox{\subfigure
{\includegraphics[scale=0.33,clip=]{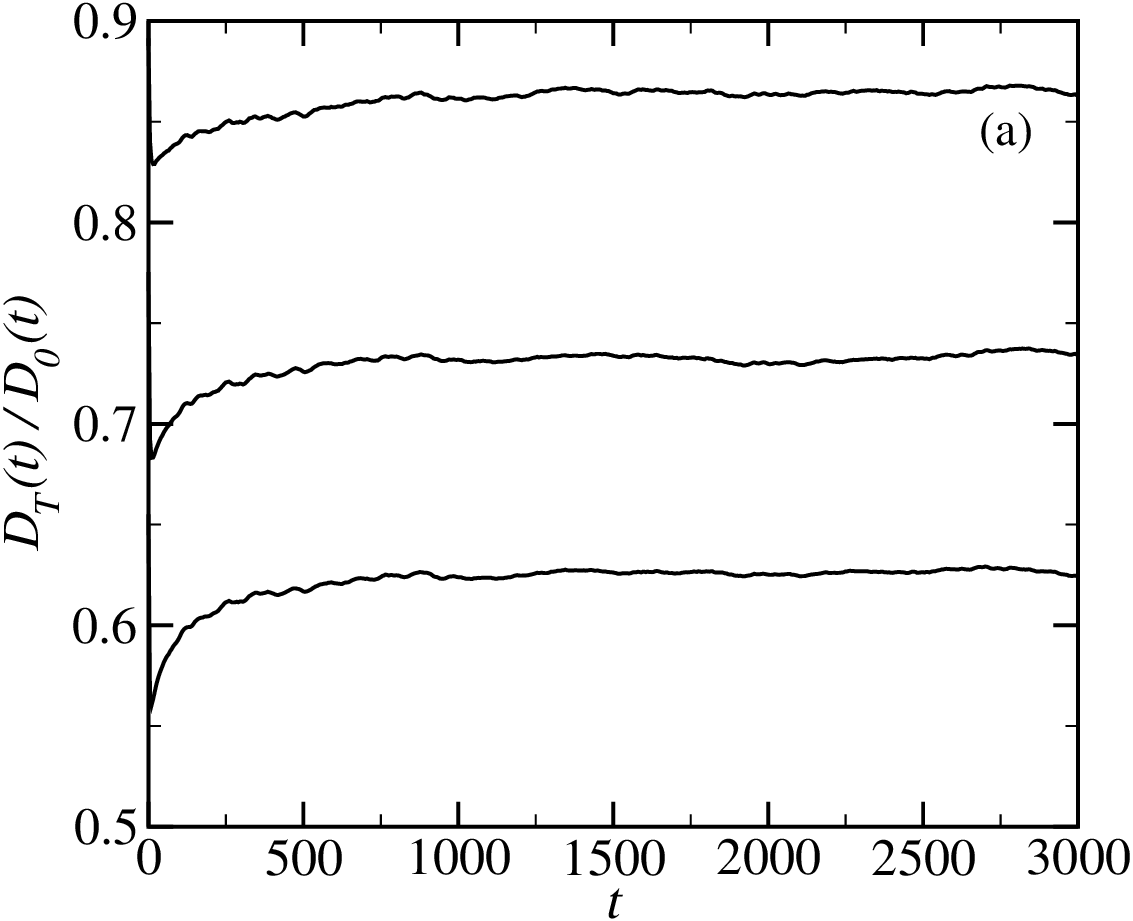}}\quad
\subfigure
{\includegraphics[scale=0.33,clip=]{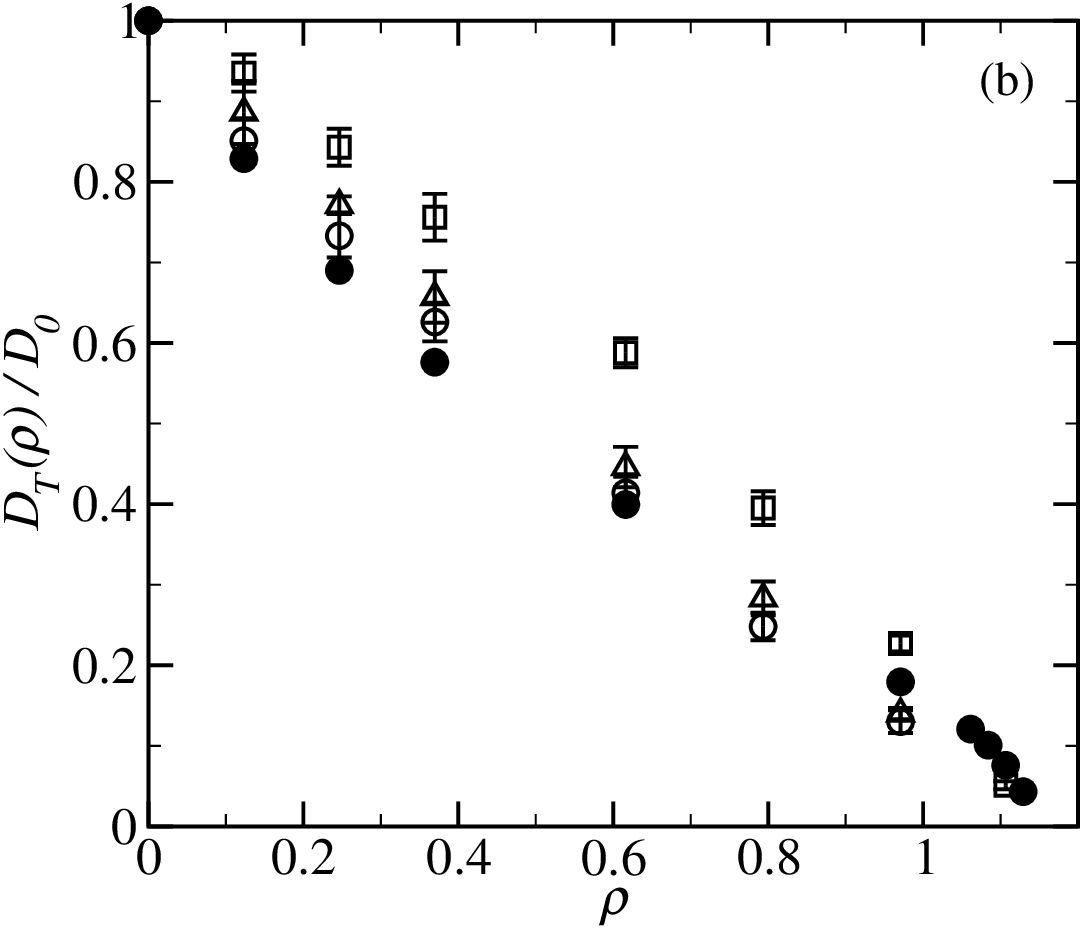}}}
\caption{\label{FIGtracer}(a) Scaled effective tracer diffusion
coefficients $D_T(\rho,t) / D_0(t)$ as a function of $t$ 
for $Sc \approx 1$. These data have been obtained from 
simulations without colloid-solvent
interactions. The colloidal densities from top to bottom
are $0.1232$, $0.2464$ and $0.3697$. (b) Scaled effective 
tracer diffusion coefficients
$D_T(\rho)/D_0$ as a function of $\rho$ 
for $Sc \approx 1$ (open circles), $Sc \approx 20$
(open triangles) and $Sc \approx 100$ (open boxes). For
reference, results without HIs \cite{lahtinen01} 
are also presented (solid circles).}
\end{figure}

We consider $D_T(\rho,t)$ normalised by the
single-particle diffusion coefficient $D_0(t) \equiv
D_T(\rho \rightarrow 0,t)$. As shown in fig.\ \ref{FIGtracer}(a), in the
limit of large times, the quantities $D_T(\rho,t)/D_0(t)$ seem to
converge to a finite limit. Although the slow logarithmic
divergences are present in our data (data not shown), the
amplitudes of the tails at late simulation times are very small
and partially masked by statistical fluctuations. In practice we
have first determined $D_0$ at a time $t_{\rm eff}$ where the
tracer particle has diffused approximately $\ell$ times its own
diameter, where $\ell = 25, 5, 2$ for $Sc = 1, 20, 100$,
respectively. The effective diffusion coefficients $D_T(\rho)$ at
finite densities have then been determined at the corresponding
time intervals.

In fig.\ \ref{FIGtracer}(b) we show a summary of our results for the
scaled $D_T(\rho) / D_0$ as a function of the density of the
disks $\rho$. For comparison we also present our earlier \acro{MD} results for
the purely dissipative case \cite{lahtinen01}. Remarkably, it is
clear that at low values of $Sc$ hydrodynamics has virtually
no influence on tracer diffusion. As $Sc$ is tuned to
larger values, we find an {\it enhancement} of $D_T(\rho ) / D_0$
which is largest at intermediate densities.
This sheds some light on the issues addressed
in ref.~\cite{Rin99}. The authors study the influence
of HIs on self-diffusion, and find that in quasi-2D systems
with {\it long-range} interactions between the colloidal particles, the
scaled diffusion coefficients are enhanced. The larger the particle density,
the stronger the enhancement \cite{Rin99}. Our data indicate that
even for {\it short-range} interactions of the form $V(r) \sim 1 / r^{n}$ with
$n = 12$, the HIs {\it may enhance} self-diffusion. Although the
mechanism leading to enhancement may be complex in a many-particle
diffusion process, we expect that it is
related to the formation of vortices, which have been shown to
enhance self-diffusion in simple liquids \cite{alder70}. Further,
although direct comparison with experiments is not possible
due to a different magnitude and range of interactions, we may
note that this result is in agreement with previous findings
\cite{Zah97,Rin99,Kol02}.

Another important transport coefficient characterizing
collective density fluctuations is the collective diffusion
coefficient $D_C(t)$, which can be defined through
a Green-Kubo relation:
\begin{equation}
D_C(t)=\xi D_{cm}(t) =
\xi \lim_{t' \rightarrow t}\frac{1}{2dNt'}\langle|\mathbf{R}
 (t')|^2\rangle .
 \label{EQgreenkubo}
\end{equation}
Here $\xi = \langle N\rangle/[\langle N^2\rangle- \langle
N\rangle^2]$ is the thermodynamic factor (proportional to the
inverse of the isothermal compressibility $\kappa_T$) and
$\mathbf{R}(t)= \sum_{i=1}^N[\mathbf{r}_i(t)-\mathbf{r}_i(0)]$ is the
center of mass (CM) displacement \cite{gomer90}.
The thermodynamic factor $\xi$,
which is a static quantity, is not affected by hydrodynamics within our 
accuracy. To estimate $\xi$ we have used the Boublik approximation
\cite{boublik75} which is in excellent agreement with \acro{MD}
simulations for most densities \cite{lahtinen01}.

In the case of dissipative hard spheres on a smooth surface, it is
an exact result that the CM mobility $D_{cm}$ is independent of
the density $\rho$ \cite{hess83}. This is because the
interparticle interactions preserve the CM momentum, and thus
$D_{cm}(\rho) = D_{cm}(0) = D_0$. However, with HIs in place this
argument does not hold any longer. In fig.\ \ref{FIGcollective}(a) we show
the scaled CM mobility as a function of density in the present
system\footnote{The CM mobilities $D_{cm}(\rho)$ have been
determined from $D_{cm}(\rho,t)$ in the same manner as the
effective self-diffusion coefficients $D_T(\rho)$ have been
obtained from $D_T(\rho,t)$.}. It is now a decreasing function of
$\rho$, and decreases much more rapidly than $D_T(\rho)/ D_0$
shown in fig.~\ref{FIGtracer}(b). It is also remarkable that, unlike
self-diffusion, $D_{cm}(\rho) / D_0$ is not sensitive to $Sc$:
while the self-diffusion of individual particles is influenced by
$Sc$, the effects on individual particles are largely independent
of each other, and thus cancel out in the CM mobility.

In fig.\ \ref{FIGcollective}(b) we show the scaled collective diffusion
coefficient that displays a {\it minimum} at small values of density.
This minimum is due to the rapid initial decay of $D_{cm}(\rho) / D_0$.
It is in striking contrast to the dissipative case
(see fig.\ \ref{FIGcollective}(b)) which is entirely determined by $\xi$.
We argue that this behaviour is generic in colloidal suspensions
governed by hydrodynamics, since then CM mobility is not constant,
but competes with $\xi$. However, the actual form of $D_C(\rho) / D_0$
may vary from one system to another, depending on the details of
interactions and the impact of the HIs.

{\it Concluding remarks} -- We have presented a detailed numerical
study of the effects of hydrodynamics on both self- and collective
diffusion of 2D repulsive colloidal particles. This has been
achieved by using a novel hybrid mesoscopic scheme for
two-component liquids. We have found that the effective tracer
diffusion coefficient $D_T(\rho)/ D_0$ is enhanced by HIs, 
an effect which becomes more pronounced with an
increasing Schmidt number. A more dramatic change has been found
in the behaviour of the effective collective diffusion coefficient
$D_C(\rho)/ D_0$, which is completely changed by hydrodynamics.
This is due to the dynamic term which now competes directly with
the thermodynamic factor, leading to non-monotonic behaviour in
contrast with the case of purely dissipative dynamics. It is an
interesting question how these results are affected when changing
from a 2D to a confined 3D geometry as in thin films, and how
effective interactions obtained from experiments \cite{Bru02}
would be manifested in the collective diffusion. Work in this
direction is in progress.

\acknowledgments
This work has been supported in part by the Academy of Finland
through its Center of Excellence Program and the National Graduate
School in Materials Physics (E.F.).

\end{document}